\documentclass[3p,times,procedia]{elsarticle}
\usepackage{nupha_ecrc}

\volume{00}

\firstpage{1}

\journalname{Nuclear Physics A}

\runauth{}

\jid{nupha}

\jnltitlelogo{Nuclear Physics A}

\usepackage{amssymb}
\biboptions{sort&compress}
\usepackage[figuresright]{rotating}
\usepackage{subcaption}
\usepackage{wrapfig}

\usepackage{lineno}
\begin{document}

\begin{frontmatter}

%% Title, authors and addresses

%% use the tnoteref command within \title for footnotes;
%% use the tnotetext command for the associated footnote;
%% use the fnref command within \author or \address for footnotes;
%% use the fntext command for the associated footnote;
%% use the corref command within \author for corresponding author footnotes;
%% use the cortext command for the associated footnote;
%% use the ead command for the email address,
%% and the form \ead[url] for the home page:
%%
%% \title{Title\tnoteref{label1}}
%% \tnotetext[label1]{}
%% \author{Name\corref{cor1}\fnref{label2}}
%% \ead{email address}
%% \ead[url]{home page}
%% \fntext[label2]{}
%% \cortext[cor1]{}
%% \address{Address\fnref{label3}}
%% \fntext[label3]{}

%% Instructions from Editor: Please use the following \dochead only in the preprint version (e-print arXiv etc.); 
%% use empty \dochead{} when submitting to Nuclear Physics A!
\dochead{XXVIIth International Conference on Ultrarelativistic Nucleus-Nucleus Collisions\\ (Quark Matter 2018)}
%\dochead{}
%% Use \dochead if there is an article header, e.g. \dochead{Short communication}
%% \dochead can also be used to include a conference title, if directed by the editors
%% e.g. \dochead{17th International Conference on Dynamical Processes in Excited States of Solids}

%%\title{Recent results from STAR Fixed-Target Collisions at Al+Au$\sqrt{s_{\textrm{NN}}}=4.9$ GeV and Au+Au$\sqrt{s_{\textrm{NN}}}=4.5$ GeV}
\title{Recent results for STAR $\sqrt{s_{NN}}=4.9$ GeV Al+Au and $\sqrt{s_{NN}}=4.5$ GeV Au+Au Fixed-Target Collisions}

%% use optional labels to link authors explicitly to addresses:
%% \author[label1,label2]{<author name>}
%% \address[label1]{<address>}
%% \address[label2]{<address>}

\author{Yang Wu for the STAR Collaboration\footnote{A list of members of the STAR collaboration and acknowledgements can be found at the end of this issue.}}

\address{Kent State University, Kent, Ohio 44242, USA}

\begin{abstract}
%% Text of abstract
We present recent results from Al(beam)+Au(target) collisions at $\sqrt{s_{NN}}=4.9$ GeV and Au+Au collisions at $\sqrt{s_{NN}}=4.5$\kern 0.16667em GeV from the STAR fixed-target (FXT) program. We report transverse mass spectra of protons, $K^0_S$ and $\Lambda$, rapidity density distributions of $\pi^\pm$, $K^0_S$ and $\Lambda$, directed flow of protons, $\pi^\pm$, $K^0_S$ and $\Lambda$, and elliptic flow of protons and $\pi^{\pm}$. These are the first measurements of pion directed and elliptic flow in this energy region. Pion and proton elliptic flow show mass ordering. Measurements are compared with published results from AGS and RHIC. These results demonstrate that the STAR detector performs well in the FXT configuration. 
\end{abstract}

\begin{keyword}
%% keywords here, in the form: keyword \sep keyword
STAR \sep fixed-target \sep FXT \sep spectra \sep directed flow \sep elliptic flow \sep strangeness \sep rapidity density
%% MSC codes here, in the form: \MSC code \sep code
%% or \MSC[2008] code \sep code (2000 is the default)

\end{keyword}

\end{frontmatter}

%%
%% Start line numbering here if you want
%%
%%\linenumbers

%% main text
\section{Introduction}
\label{}

%% inserted figures by Yang
\begin{wrapfigure}{r}{0.40\textwidth} % was 0.35
\vspace{-40pt}
\begin{center}
\includegraphics[width=0.3\textwidth]  {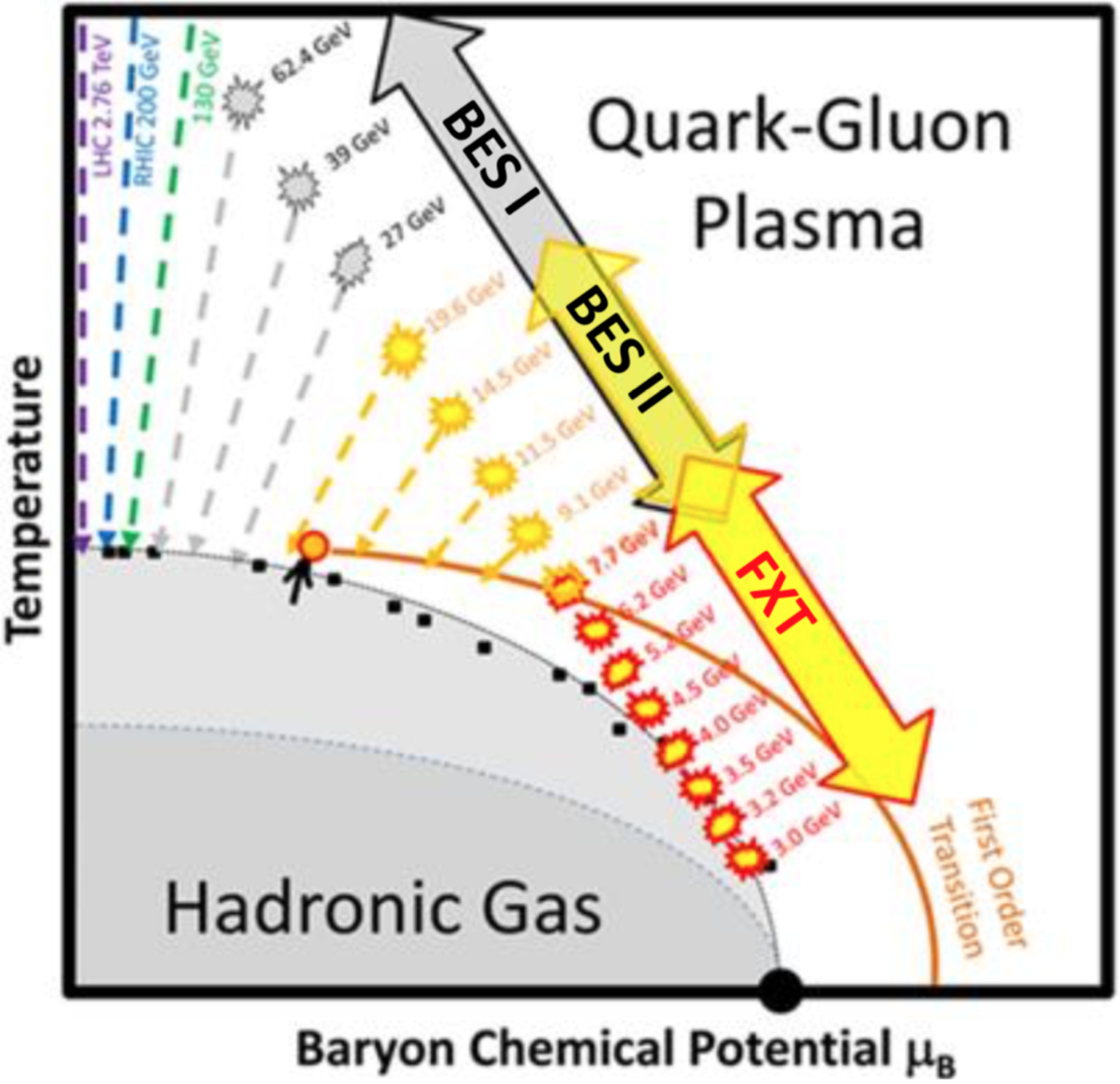}  % was 0.32
\end{center}
\vspace{-15pt}
\caption{Schematic QCD phase diagram, illustrating the conjectured critical point, phase transition lines, and the range of the FXT program and beam energy scan in collider mode.}
\vspace{-25pt}  % was -12
%\label{fig:Phase_diagram_cartoon}
\end{wrapfigure}

The goals of the STAR beam energy scan (BES) program include searches for a possible QCD critical point, and for the turn-off of signatures of quark-gluon plasma (QGP); study of the transition between hadronic and partonic matter is also of great interest \cite{Aggarwal:2010cw}. RHIC BES phase I (BES-I) has yielded interesting results below $\sqrt{s_{NN}} = 19.6$\kern 0.16667em GeV in azimuthal anisotropy for identified hadrons, kaon to pion ratios, and net-proton higher moments. These features continue to the lowest RHIC collider-mode energy, $\sqrt{s_{NN}}=7.7$ GeV, and motivate investigation at even lower energies. The STAR fixed-target (FXT) program \cite{Kathryn:2017qm} extends the energy reach from $\sqrt{s_{NN}}=7.7$ GeV to $\sqrt{s_{NN}}$ 

%% inserted figures by Yang
\begin{wrapfigure}{r}{0.47\textwidth}  % was 0.4
\vspace{-20pt} % was -80
\begin{center}
\includegraphics[width=0.47\textwidth]{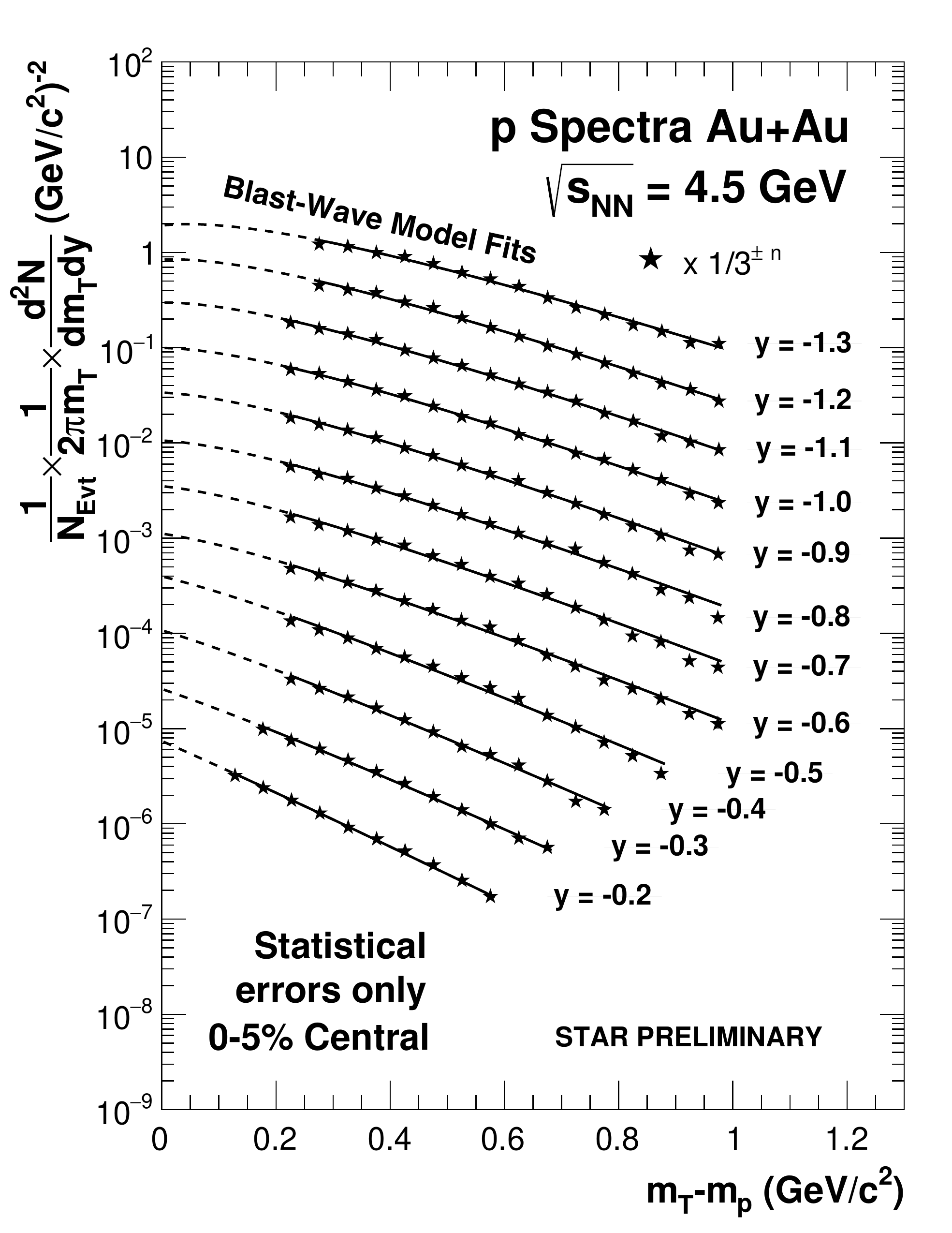} % was 0.4
\end{center}
\vspace{-15pt}
\caption{Proton transverse mass spectra in 4.5 GeV Au+Au collisions.}
\label{fig:Proton_dNdmT_4_5}
\vspace{-20pt}
\end{wrapfigure}

\noindent = 3.0 GeV, corresponding to baryon chemical potentials from 420 MeV to about 720 MeV. The directed flow ($v_1$) for protons and $\Lambda$ hyperons at 7.7 GeV and above \cite{PhysRevLett.112.162301, PhysRevLett.120.062301} suggests a qualitative resemblance to the ``softest point" of the equation of state predicted by hydrodynamic models. 

Net-proton higher moment measurements from energies below 7.7 GeV could help determine the type of phase transition or reveal evidence for critical fluctuations \cite{Luo:2017faz}. However, RHIC cannot operate in collider mode below 7 GeV, due to steeply decreasing luminosity. By inserting a target into the beam pipe and circulating one beam in RHIC, we can study FXT collisions below $\sqrt{s_{NN}}=7$ GeV. Figure 1 shows an example of a schematic phase diagram, and illustrates the possible region probed by FXT measurements. 

During a brief test in 2015, STAR collected approximately 1.3 million FXT events with centrality 0-30\%.

\section{Results}
\subsection{STAR FXT Particle Yield Results}

Figure \ref{fig:Proton_dNdmT_4_5} presents proton spectra for 0-5\% centrality in Au+Au collisions at $\sqrt{s_{NN}}=4.5$\kern 0.16667em GeV as a function of transverse mass $m_T-m_p$ in several rapidity bins, each with a width $\Delta y=0.1$. These spectra were fitted using the Blast-Wave model, after correcting for detector efficiency, acceptance, energy loss and hadronic background. Overall, fits describe the data across the STAR FXT acceptance range.

%% inserted figures by Yang
\begin{figure}[ht]
\centering
\begin{subfigure}[b]{0.425\textwidth} % was 0.403
\centering\includegraphics[width=\textwidth]{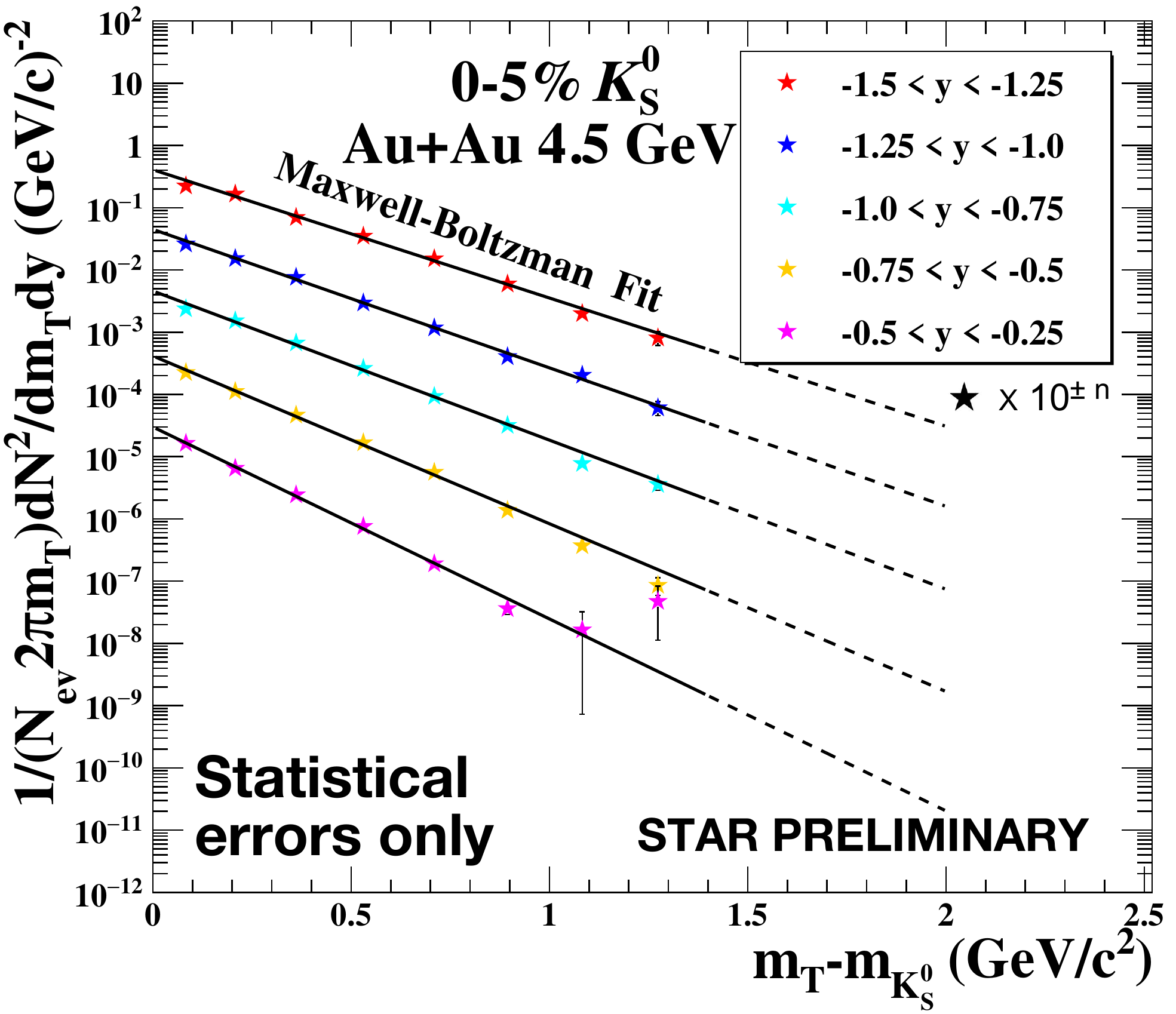}
\end{subfigure}\hspace{5mm}
\begin{subfigure}[b]{0.475\textwidth} % was 0.397
\centering\includegraphics[width=\textwidth]{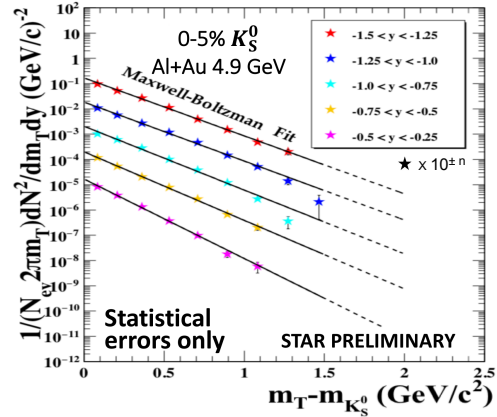}
\end{subfigure}
\caption{$K^0_S$ transverse mass spectra for central $\sqrt{s_{NN}}=4.5$ GeV Au+Au and 4.9 GeV Al+Au collisions.}
\vspace{-10pt}
\label{fig:Kaon_spectra}
\end{figure}

\begin{figure}[ht]
\centering
%\begin{subfigure}[b]{0.32\textwidth}
%\centering\includegraphics[width=\textwidth]{Figures/Lambda_dNdmT_4_5.png}
%\end{subfigure}
\begin{subfigure}[b]{0.425\textwidth} % was 0.382
\centering\includegraphics[width=\textwidth]{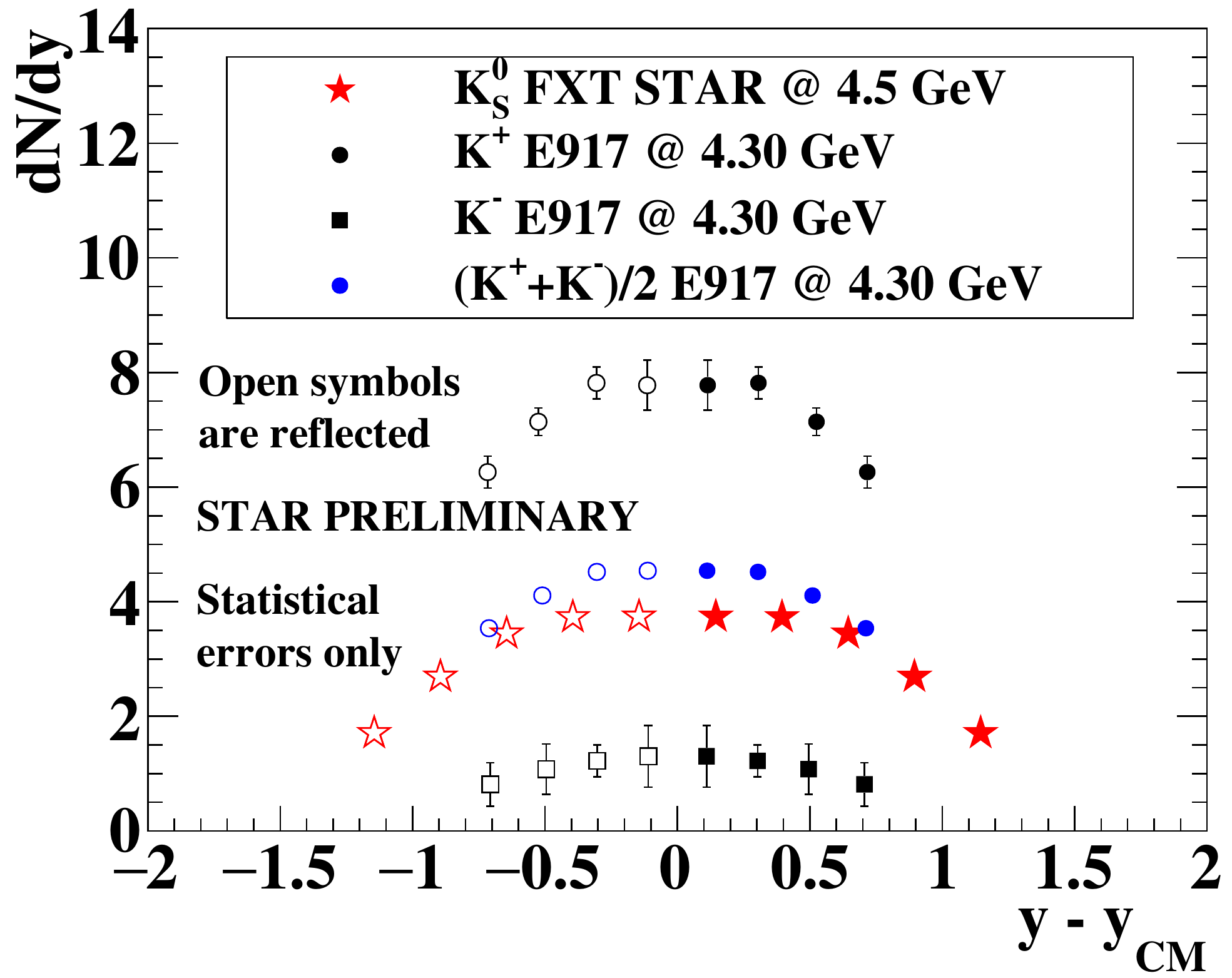}
\end{subfigure}\hspace{5mm}
\begin{subfigure}[b]{0.475\textwidth} % was 0.418
\centering\includegraphics[width=\textwidth]{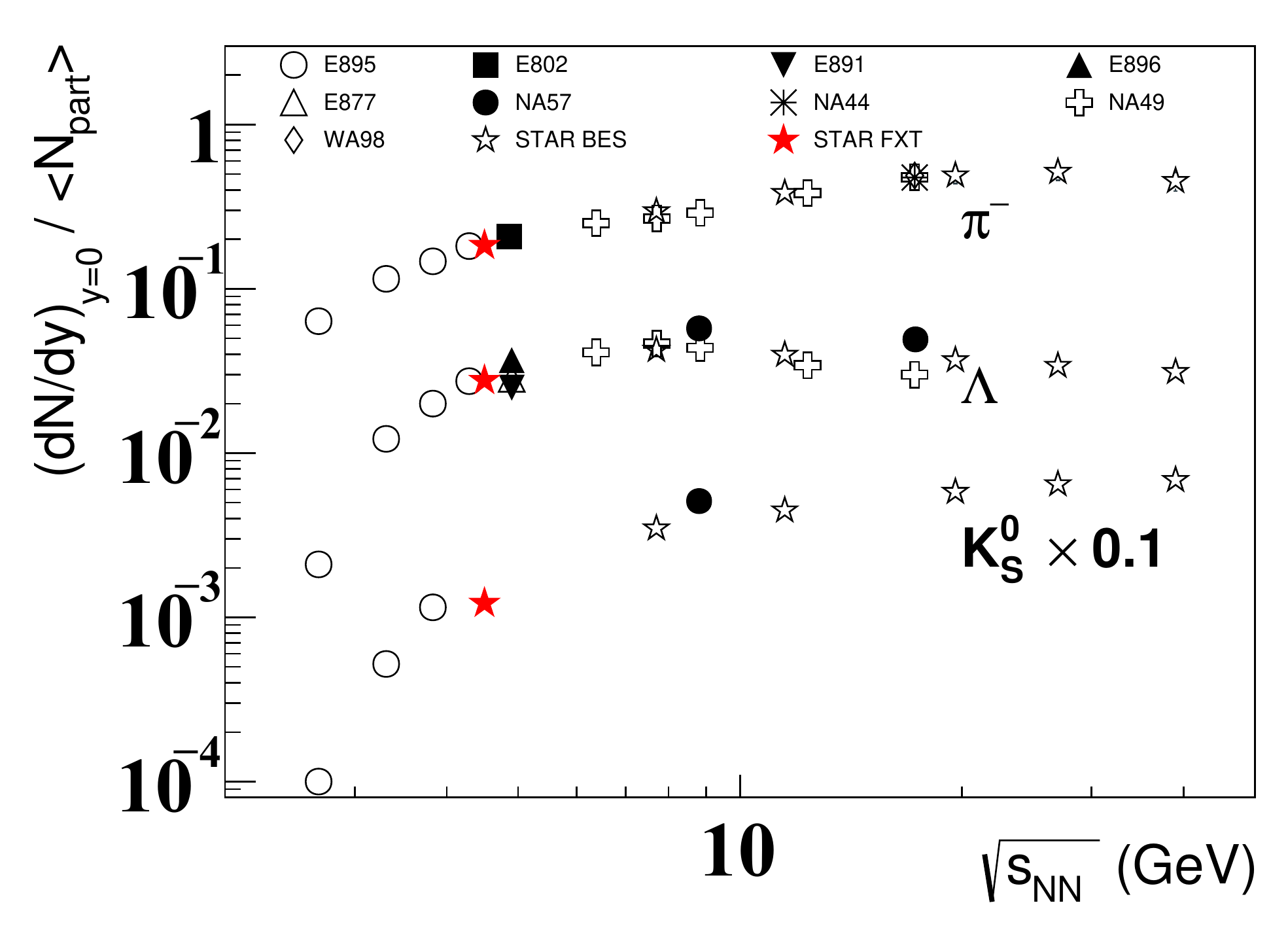}
\end{subfigure}
\caption{(Left) Kaon $dN/dy$ for Au+Au at or near 4.5 GeV. (Right) Peak $dN/dy$ vs. beam energy for $\pi^-$, $\Lambda$ and $K^0_S$.}
\vspace{-12pt}
\label{fig:K_Lambda_dNdy}
\end{figure}

In Fig. \ref{fig:Kaon_spectra}, $m_T$ spectra of $K^0_S$ from 4.5 GeV Au+Au and 4.9 GeV Al+Au collisions are shown for 0-5\% centrality in several rapidity bins of width $\Delta y=0.25$. These spectra are fitted using a Maxwell-Boltzmann functional form. The spectra and fits for $\Lambda$ (not shown) are found to be similar to Fig. \ref{fig:Kaon_spectra} in their range and fit quality. Kaon rapidity densities, $dN/dy$, are extracted from the fits and compared with AGS results in Fig. \ref{fig:K_Lambda_dNdy} (left). The STAR $K^0_S$ $dN/dy$ falls between E917 $K^+$ and $K^-$ results and are slightly lower than $(K^+ + K^-)/2$ from E917\kern 0.16667em \cite{Ogilvie:1998ru}. The amplitudes and widths of the current $K^0_S$ $dN/dy$ indicate good agreement between STAR FXT and published AGS results.

%% inserted figures by Yang

In Fig. \ref{fig:K_Lambda_dNdy} (right), $dN/dy$ distributions of $\pi^-$, $K^0_S$ and $\Lambda$ obtained in Au+Au collisions at $\sqrt{s_{NN}}=4.5$\kern 0.16667em GeV near mid-rapidity, normalized by mean participant number $N_{\rm part}$, are compared across various experiments as a function of beam energy. The current STAR FXT rapidity densities are consistent with the trends defined by prior measurements at higher and lower energies \cite{PhysRevLett.88.062301,PhysRevC.66.044907,0954-3899-30-8-008,PhysRevLett.93.022302,PhysRevC.67.014906,PhysRevC.68.054905,PINKENBURG2002495,AKIBA1996139,PhysRevC.63.014902,AHMAD199635}.
% * <ywu27@kent.edu> 2018-06-28T04:19:35.702Z:
%
% ^.

\subsection{STAR FXT Flow Results}
\begin{figure}[ht]
\centering
\begin{subfigure}[b]{0.475\textwidth} % was 0.395
\centering\includegraphics[width=\textwidth]{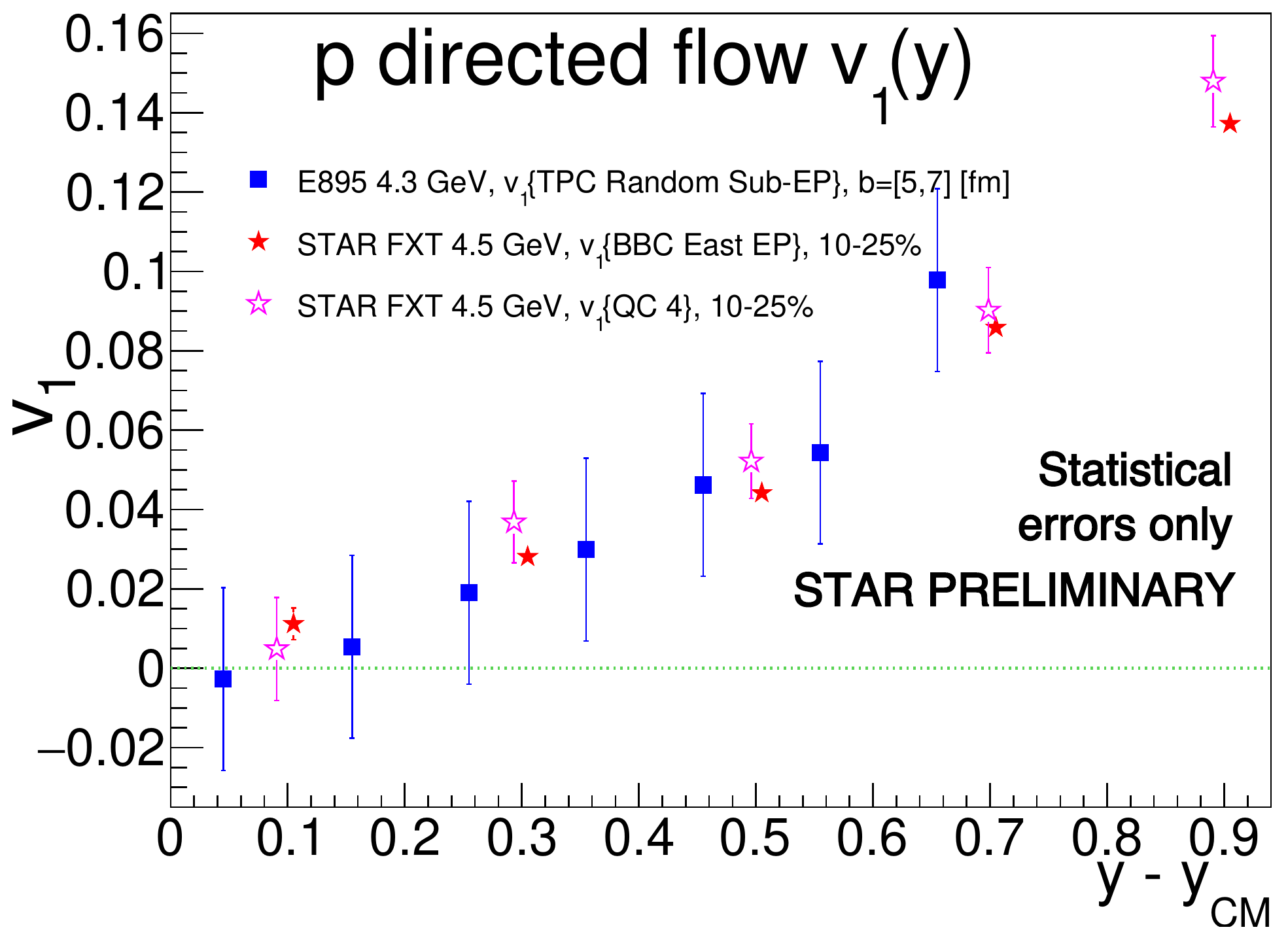}
\end{subfigure}\hspace{5mm}
\begin{subfigure}[b]{0.475\textwidth} % was 0.405
\centering\includegraphics[width=\textwidth]{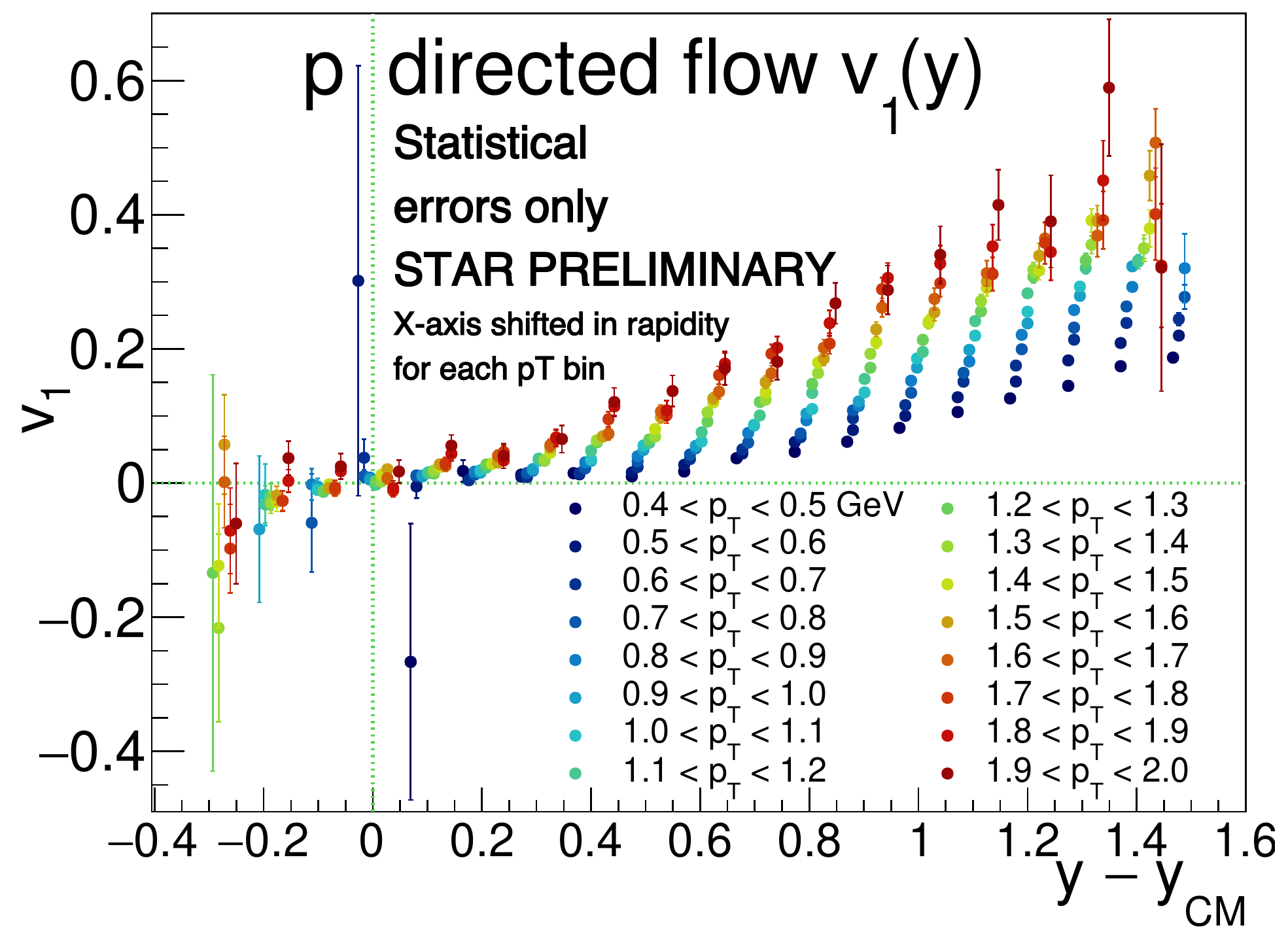}
\end{subfigure}
\caption{(Left) proton $v_1(y)$ for FXT 4.5 GeV Au+Au, based on two separate methods \cite{Poskanzer:1998yz,Bilandzic:2010jr}, and compared with E895 \cite{PhysRevLett.84.5488}.\\ (Right) The same FXT proton $v_1(y)$ in narrow intervals of transverse momentum.}
\vspace{10pt}
\label{fig:Proton_v1y_4_5}
\end{figure}

The left panel of Fig. \ref{fig:Proton_v1y_4_5} presents proton $v_1(y)$ from 4.5 GeV Au+Au (red markers). E895 data for 4.3 GeV Au+Au at similar $p_T$ and centrality are also plotted \cite{PhysRevLett.84.5488}. STAR FXT results are consistent with E895 but have much smaller statistical errors. In the right panel of Fig. \ref{fig:Proton_v1y_4_5}, proton $v_1(y)$ is plotted in narrow intervals of $p_T$. The proton acceptance in FXT mode at this beam energy, with the standard selection of $0.4 < p_T < 2.0$ GeV/$c$ used in prior studies \cite{PhysRevLett.112.162301, PhysRevLett.120.062301, PhysRevLett.84.5488}, was found to have
negligible impact on proton $v_1(y)$ down to $y - y_{cm} =0$. 
%% Except for some data points limited by statistics, the STAR FXT detector acceptance has small to negligible effects on proton directed flow analysis. 

%% inserted figures by Yang
\vspace{10pt}

%% inserted figures by Yang
\begin{figure}[!htb]
\centering
\begin{subfigure}[b]{0.5\textwidth}  % was 0.395
\centering\includegraphics[width=\textwidth]{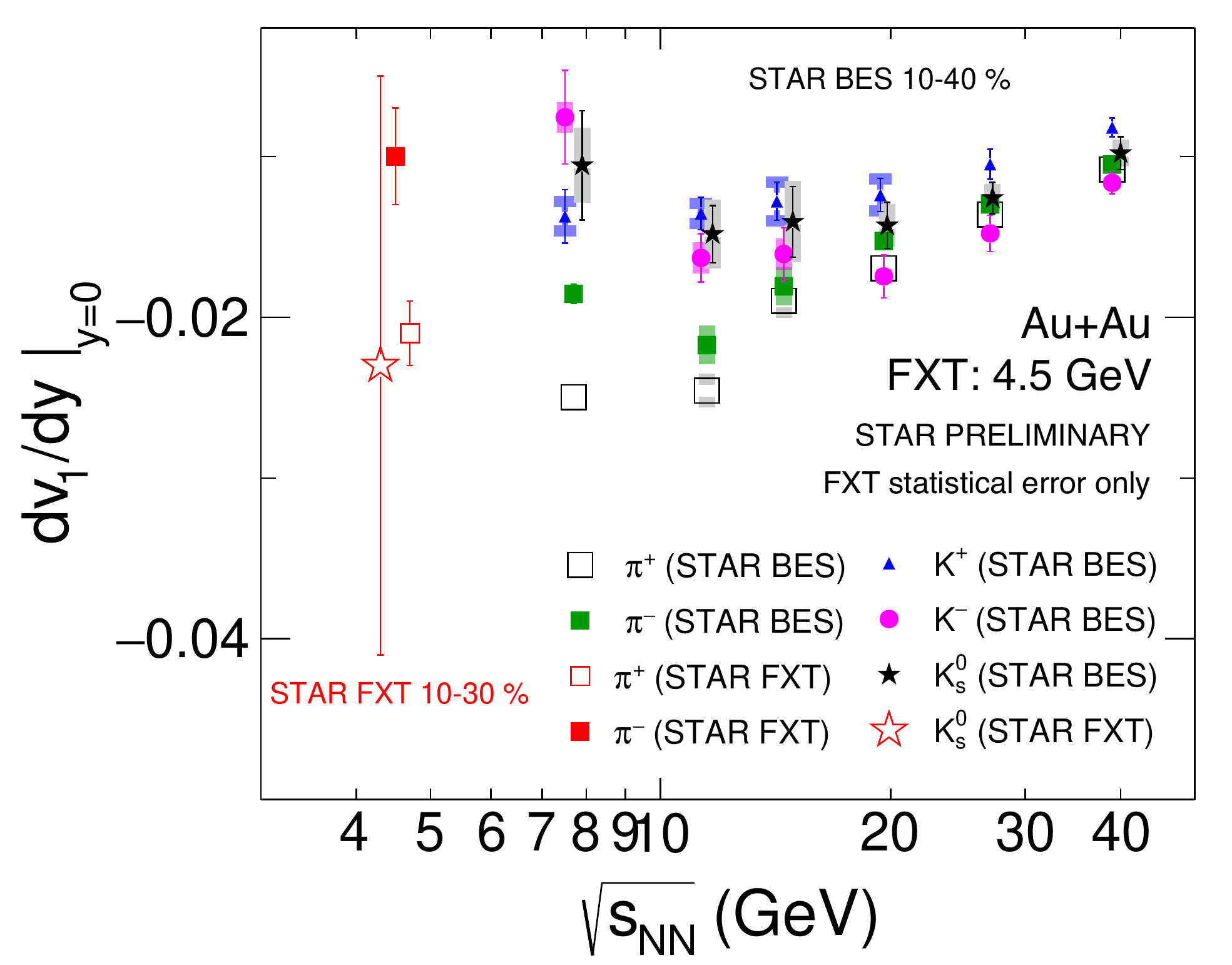}
\end{subfigure}%\hspace{5mm}
\begin{subfigure}[b]{0.5\textwidth}  % was 0.405
\centering\includegraphics[width=\textwidth]{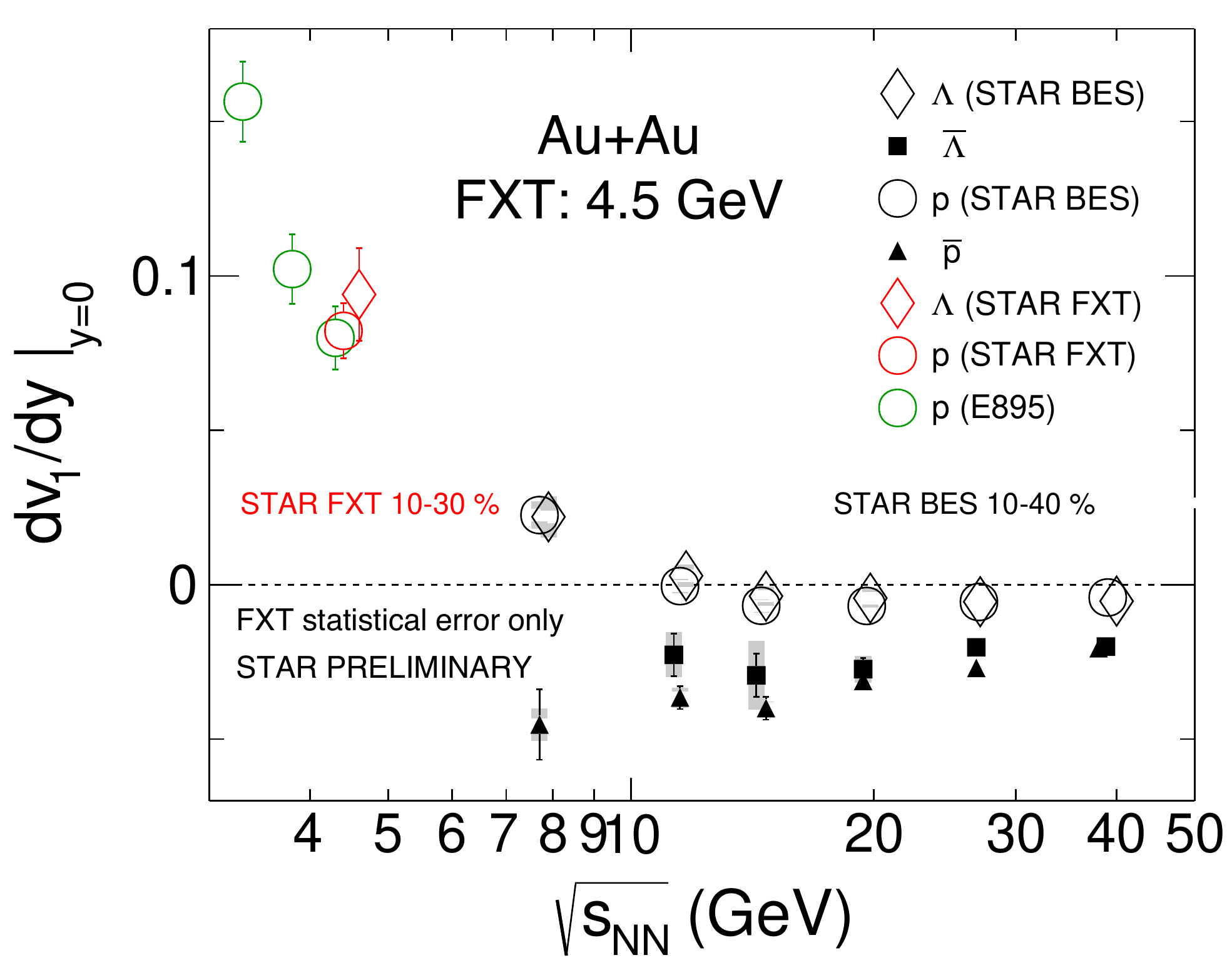}
\end{subfigure}
\caption{Mid-rapidity directed flow slope versus beam energy for mesons (left panel) and for baryons (right panel).}
\vspace{-15pt}
\label{fig:v1_slopes_4_5}
\end{figure}

FXT directed flow slope $dv_1/dy$ near midrapidity for protons, $\pi^{\pm}$, $K^0_S$ and $\Lambda$ at 4.5 GeV are compared in the left panel of Fig. \ref{fig:v1_slopes_4_5} with STAR collider-mode results \cite{PhysRevLett.112.162301, PhysRevLett.120.062301} and with E895 \cite{PhysRevLett.84.5488}. The STAR FXT $\pi^{\pm}$ and $K^0_S$ measurements continue the trend of negative $dv_1/dy$ for mesons observed at 7.7 GeV and above. The STAR FXT $dv_1/dy$ for protons at 4.5 GeV is in good agreement with E895 at 4.3 GeV, and both are consistent with a smooth interpolation between higher and lower energies. 

\begin{wrapfigure}{r}{0.45\textwidth} % was 0.42
\vspace{-15pt}
\begin{center}
\includegraphics[width=0.46\textwidth] {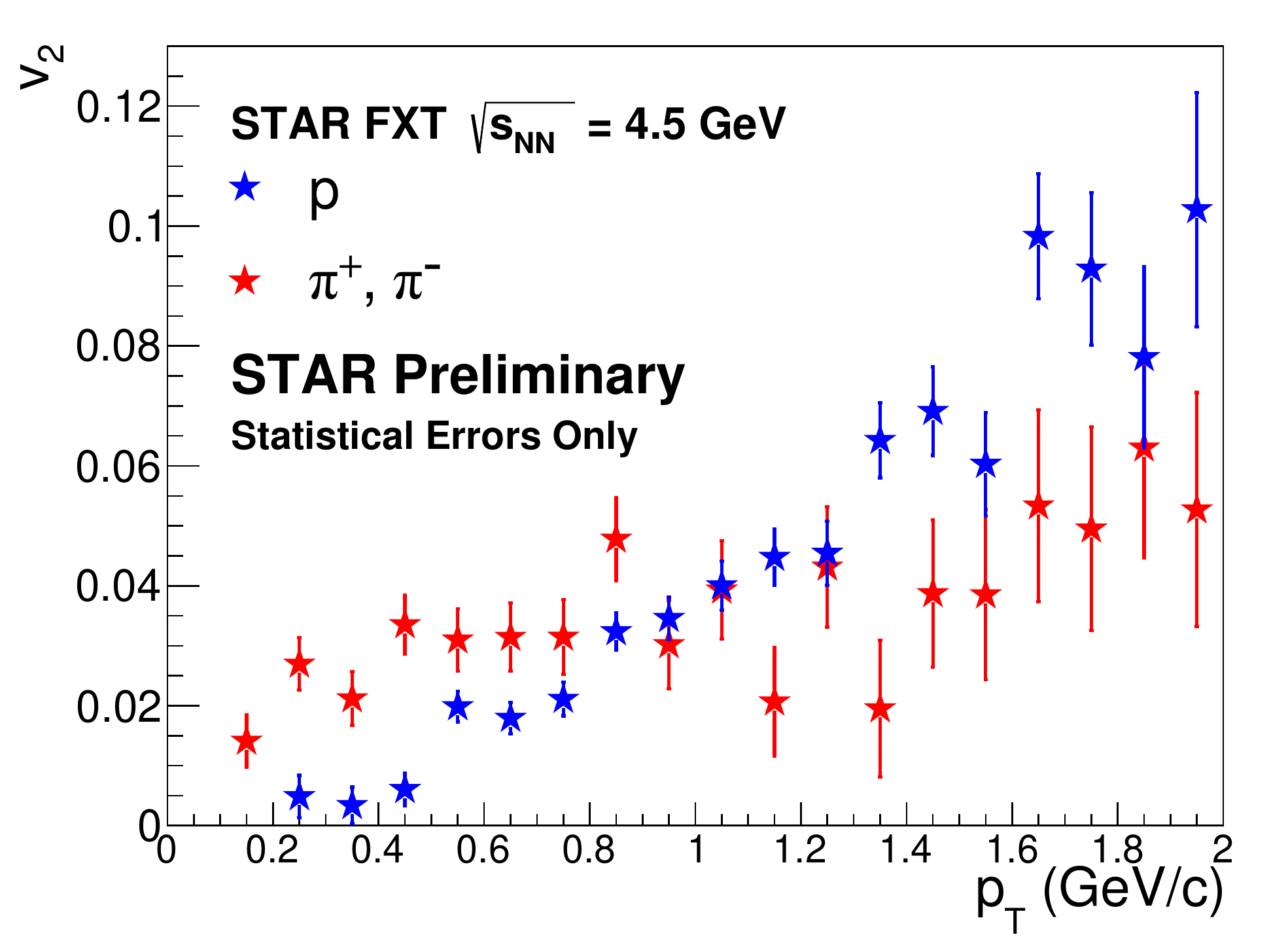} % was 0.42
\end{center}
\vspace{-15pt}
\caption{4.5 GeV Au+Au $v_2(p_T)$ for $\pi^\pm$ and protons.}
\vspace{-5pt}
\label{fig:v2ofpT}
\end{wrapfigure}

Figure \ref{fig:v2ofpT} presents FXT 4.5 GeV Au+Au proton and pion elliptic flow $v_2(p_T)$. The observed mass ordering of proton and pion $v_2$ resembles measurements at higher energies \cite{Adams:2005dq, Adamczyk:2013gw, Adamczyk:2015fum}. 
%% We have first measurements of pion directed and elliptic flows at this energy range. (Already stated in the abstract, so save space by not repeating.)

\vspace{15pt}
\section{Future FXT Plans}

%% inserted figures by Yang

In future runs, STAR will have multiple detector upgrades \cite{QianYang:2018qm}, including the Inner Time Projection Chamber (iTPC), Endcap Time Of Flight (eTOF) and Event Plane Detector (EPD). The EPD is already in operation. These upgrades extend the detector acceptance, and improve the particle identification capability, event plane resolution and centrality determination. Running in FXT mode for two days at each of several energies between $\sqrt{s_{NN}}=3.0$ GeV and 7.7 GeV will permit acquisition of $\sim$100 million events per energy and will extend the reach of baryon chemical potential to about 720 MeV. The highest FXT energy will overlap with the lowest collider-mode energy, allowing detailed cross checks.

%% The Appendices part is started with the command \appendix;
%% appendix sections are then done as normal sections
%% \appendix

%% \section{}
%% \label{}

%% References
%%
%% Following citation commands can be used in the body text:
%% Usage of \cite is as follows:
%%   \cite{key}         ==>>  [#]
%%   \cite[chap. 2]{key} ==>> [#, chap. 2]
%%

%% References with BibTeX database:
\vspace{-10pt}

\bibliographystyle{elsarticle-num}
%\bibliography{proceeding.bib}

%% Authors are advised to use a BibTeX database file for their reference list.
%% The provided style file elsarticle-num.bst formats references in the required Procedia style

%% For references without a BibTeX database:

% \begin{thebibliography}{00}

%% \bibitem must have the following form:
%%   \bibitem{key}...
%%

% \bibitem{}

% \end{thebibliography}

\end{document}